\documentclass{emulateapj}

\newcommand{\pks}{PKS~B2322$-$275}
\newcommand{\refsrc}{PKS~B2331$-$312}
\newcommand{\pksb}{PKS~B2318$-$195}
\newcommand{\kms}{\ensuremath{{\rm km\,s}^{-1}}}
\newcommand{\pasa}{PASA}

\def\ASU{{11}}
\def\ANU{{12}}
\def\CASS{{5}}
\def\Curtin{{2}}
\def\CfA{{10}}
\def\Haystack{{13}}
\def\MIT{{7}}
\def\RRI{{14}}
\def\SKASA{{8}}
\def\RU{{9}}
\def\Tata{{4}}
\def\UMelbourne{{18}}
\def\USydney{{6}}
\def\UW{{16}}
\def\UWisc{{1}}
\def\Toronto{15}
\def\Victoria{{17}}
\def\CAASTRO{{3}}

\slugcomment{ApJL, in press}

\shorttitle{MWA observations of Interplanetary Scintillation}
\shortauthors{Kaplan et al.}

\begin{document}

\title{Murchison Widefield Array Observations of Anomalous
  Variability: A Serendipitous Night-time Detection of Interplanetary Scintillation}

 \author{D.~L.~Kaplan\altaffilmark{\UWisc},
  S.~J.~Tingay\altaffilmark{\Curtin,\CAASTRO},
P.~K.~Manoharan\altaffilmark{\Tata},
  J.-P.~Macquart\altaffilmark{\Curtin,\CAASTRO}, 
P. Hancock\altaffilmark{\Curtin,\CAASTRO},
  J.~Morgan\altaffilmark{\Curtin},
  D.~A.~Mitchell\altaffilmark{\CASS,\CAASTRO}, 
R.~D.~Ekers\altaffilmark{\CASS},
  R.~B.~Wayth\altaffilmark{\Curtin,\CAASTRO}, 
C.~Trott\altaffilmark{\Curtin,\CAASTRO},
T.~Murphy\altaffilmark{\USydney,\CAASTRO},
D.~Oberoi\altaffilmark{\Tata}, 
I.~H.~Cairns\altaffilmark{\USydney},
L.~Feng\altaffilmark{\MIT}, 
N.~Kudryavtseva\altaffilmark{\Curtin},
  G.~Bernardi\altaffilmark{\SKASA,\RU,\CfA}, 
J.~D.~Bowman\altaffilmark{\ASU}, 
F.~Briggs\altaffilmark{\ANU},
  R.~J.~Cappallo\altaffilmark{\Haystack}, 
A.~A.~Deshpande\altaffilmark{\RRI},
  B.~M.~Gaensler\altaffilmark{\USydney,\CAASTRO,\Toronto},
L.~J.~Greenhill\altaffilmark{\CfA},
N.~Hurley-Walker\altaffilmark{\Curtin},
  B.~J.~Hazelton\altaffilmark{\UW}, 
M.~Johnston-Hollitt\altaffilmark{\Victoria},
  C.~J.~Lonsdale\altaffilmark{\Haystack}, 
S.~R.~McWhirter\altaffilmark{\Haystack}, M.~F.~Morales\altaffilmark{\UW},
  E.~Morgan\altaffilmark{\MIT}, 
S.~M.~Ord\altaffilmark{\Curtin,\CAASTRO},
  T.~Prabu\altaffilmark{\RRI}, 
N.~Udaya~Shankar\altaffilmark{\RRI},
 K.~S.~Srivani\altaffilmark{\RRI},
  R.~Subrahmanyan\altaffilmark{\RRI,\CAASTRO}, 
  R.~L.~Webster\altaffilmark{\UMelbourne,\CAASTRO}, 
A.~Williams\altaffilmark{\Curtin},
  and C.~L.~Williams\altaffilmark{\MIT}
}

\altaffiltext{\UWisc}{Department of Physics, University of
  Wisconsin--Milwaukee, Milwaukee, WI 53201, USA; \email{kaplan@uwm.edu}}
\altaffiltext{\Curtin}{International Centre for Radio Astronomy Research, Curtin University, Bentley, WA 6102, Australia}
\altaffiltext{\CAASTRO}{ARC Centre of Excellence for All-sky
  Astrophysics (CAASTRO)}
\altaffiltext{\Tata}{Radio Astronomy Centre, National Centre for Radio Astrophysics, Tata
Institute of Fundamental Research, Ooty 643001, India}
\altaffiltext{\CASS}{CSIRO Astronomy and Space Science (CASS), PO Box
  76, Epping, NSW 1710, Australia}
\altaffiltext{\USydney}{Sydney Institute for Astronomy, School of Physics, The University of Sydney, NSW 2006, Australia}
\altaffiltext{\MIT}{Kavli Institute for Astrophysics and Space
  Research, Massachusetts Institute of Technology, Cambridge, MA
  02139, USA}
\altaffiltext{\SKASA}{SKA SA, 3rd Floor, The Park, Park Road, Pinelands, 7405, South Africa}
\altaffiltext{\RU}{Department of Physics and Electronics, Rhodes University, PO Box 94, Grahamstown 6140, South Africa}
\altaffiltext{\CfA}{Harvard-Smithsonian Center for Astrophysics, Cambridge, MA 02138, USA}
\altaffiltext{\ASU}{School of Earth and Space Exploration, Arizona
  State University, Tempe, AZ 85287, USA}
\altaffiltext{\ANU}{Research School of Astronomy and Astrophysics, Australian National University, Canberra, ACT 2611, Australia}
\altaffiltext{\Haystack}{MIT Haystack Observatory, Westford, MA 01886, USA}
\altaffiltext{\RRI}{Raman Research Institute, Bangalore 560080, India}
\altaffiltext{\Toronto}{Dunlap Institute for Astronomy and Astrophysics,
 University of Toronto, 50 St.\ George Street, Toronto, ON M5S 3H4, Canada}
\altaffiltext{\UW}{Department of Physics, University of Washington, Seattle, WA 98195, USA}
\altaffiltext{\Victoria}{School of Chemical \& Physical Sciences,
  Victoria University of Wellington, Wellington 6140, New Zealand}
\altaffiltext{\UMelbourne}{School of Physics, The University of Melbourne, Parkville, VIC 3010, Australia}

\begin{abstract}
We present observations of high-amplitude rapid (2\,s) variability
toward two bright, compact extragalactic radio sources out of several
hundred of the brightest radio sources in one of the $30\degr\times
30\degr$ MWA Epoch of Reionization fields using the Murchison
Widefield Array (MWA) at 155\,MHz.  After rejecting intrinsic,
instrumental, and ionospheric origins we consider the most likely
explanation for this variability to be interplanetary scintillation
(IPS), likely the result of a large coronal mass ejection propagating
from the Sun.  This is confirmed by roughly contemporaneous
observations with the Ooty Radio Telescope.  We see evidence for
structure on spatial scales ranging from $<$1000\,km to
$>10^6$\,km. The serendipitous night-time nature of these detections
illustrates the new regime that the MWA has opened for IPS studies
with sensitive night-time, wide-field, low-frequency observations.
This regime complements traditional dedicated strategies for observing
IPS and can be utilized in real-time to facilitate dedicated follow-up
observations.  At the same time, it allows large-scale surveys for
compact (arcsec) structures in low-frequency radio sources despite the
$2\arcmin$ resolution of the array.

\end{abstract}

\section{Introduction}
Interplanetary scintillation (IPS) --- the rapid ($<$1\,s--$10\,$s)
flux density variability of distant compact radio sources due to
radio-wave propagation through inhomogeneities in the ionized solar
wind --- was discovered by \citet{Clarke64} and published by
\citet{1964Natur.203.1214H}; the discovery led to high-time-resolution
instrumentation that, in turn, led to the discovery of pulsars
\citep{1968Natur.217..709H}.  Interplanetary scintillation is an important technique in
monitoring the structure and evolution of the solar wind
\citep[e.g.,][]{1978SSRv...21..411C,1990MNRAS.244..691M,1998JGR...10312049J},
particularly as a probe of major perturbations caused by solar flares
and Coronal Mass Ejections (CMEs)
\citep[e.g.,][]{2003JGRA..108.1220T,2010SoPh..265..137M}, and in
constraining the nature of background radio sources
\citep{1968MNRAS.138..393L}.

Observations of IPS are typically made at low radio-frequencies,
$<400\,$MHz, where the effects of plasma inhomogeneities are most
apparent.  At facilities such as the Ooty radio telescope
\citep{2010SoPh..265..137M}, the Mexican Array Radio Telescope
(MEXART; \citealt{2010SoPh..265..309M}), and the Solar Wind Imaging
Facility, Toyokawa (SWIFT;
\citealt{2011RaSc...46.0F02T})/Solar-Terrestrial Environment
Laboratory (STEL; \citealt{1990SSRv...53..173K}),
observations measure the modulation of hundreds of pre-selected radio
sources each day and follow solar-wind
events identified elsewhere.  The new generation of low-frequency,
wide-field radio telescopes such as the Murchison Widefield Array
(MWA; \citealt{lon09,tin13}) and Low Frequency Array  \citep[LOFAR; ][]{2013AA...556A...2V} can
undertake both dedicated observations of IPS
\citep{2010SoPh..265..293O,2011AGUFMSH31C2020B,2013SoPh..285..127F}
--- often at high time resolution --- and make serendipitous
discoveries, as discussed here.  These ground-based radio IPS
facilities complement space-based heliospheric imagers such as STEREO
\citep{2009SoPh..254..387E}, SDO \citep{2012SoPh..275..229S}, and
Hinode \citep{2007SoPh..243...19C}, and in-situ solar wind
measurements from other spacecraft.

The MWA is a low frequency (80--300\,MHz) interferometer located in
Western Australia, with solar, heliospheric, and ionospheric studies
as one of its major focuses \citep{bow13}.  The advantages of the MWA
for IPS
are: its southern hemisphere location (all  major IPS facilities
are in the northern hemisphere); its very wide field of view (over
$600\,{\rm deg}^2$ at 150\,MHz); its high sensitivity for short
integrations ($\sim$100\,mJy/beam rms for a 1\,s integration); and its
capability to make sub-second flux density measurements.
Here we
present the first serendipitous, exploratory observations of IPS using
the MWA in its normal imaging mode with 2\,s time resolution.

\section{Observations and Data Analysis}
\label{sec:obs}
Data for this investigation were from commensal MWA
proposals\footnote{http://www.mwatelescope.org/astronomers} G0009
(``Epoch of Reionisation'') and G0005 (``Search for Variable and
Transient Sources in the EoR Fields with the MWA'').  The observations
are of one of the MWA Epoch of Reionisation (EoR) fields (designated
EoR0, and centered at J2000 RA=$0^{\rm h}$, Dec=$-27\degr$; see
Figure~\ref{fig:image}) and occurred between 11:30 and 14:14 UT on
2014 November 06, while the field was within approximately two hours
of transit.  The observations ({see \citealt{loi15} for
more details}) used 40\,kHz fine channels across a
30.72\,MHz bandwidth centered at 155\,MHz and 2\,s integrations; the
correlated data were written to files of 2\,min each, with gaps of
16--24\,s between adjacent files.

Images were searched for Fast Radio Bursts \citep{lor07,tho13} using a
processing pipeline which will be described elsewhere (Tingay et al.,
in preparation).  A by-product of this processing is the total power
variability for every object in the MWA field of view at 2\,s cadence.
Very strong detections were triggered at the position of \pks\ (a
redshift 1.27 quasar; \citealt{1996ApJS..107...19M}) during a two
minute period (13:30:00--13:32:00\,UT).  This variability prompted the
detailed follow-up analysis described below.

\subsection{Real-Time System analysis}
\begin{figure*}
% \plotone{pks2322_image.pdf}
\plotone{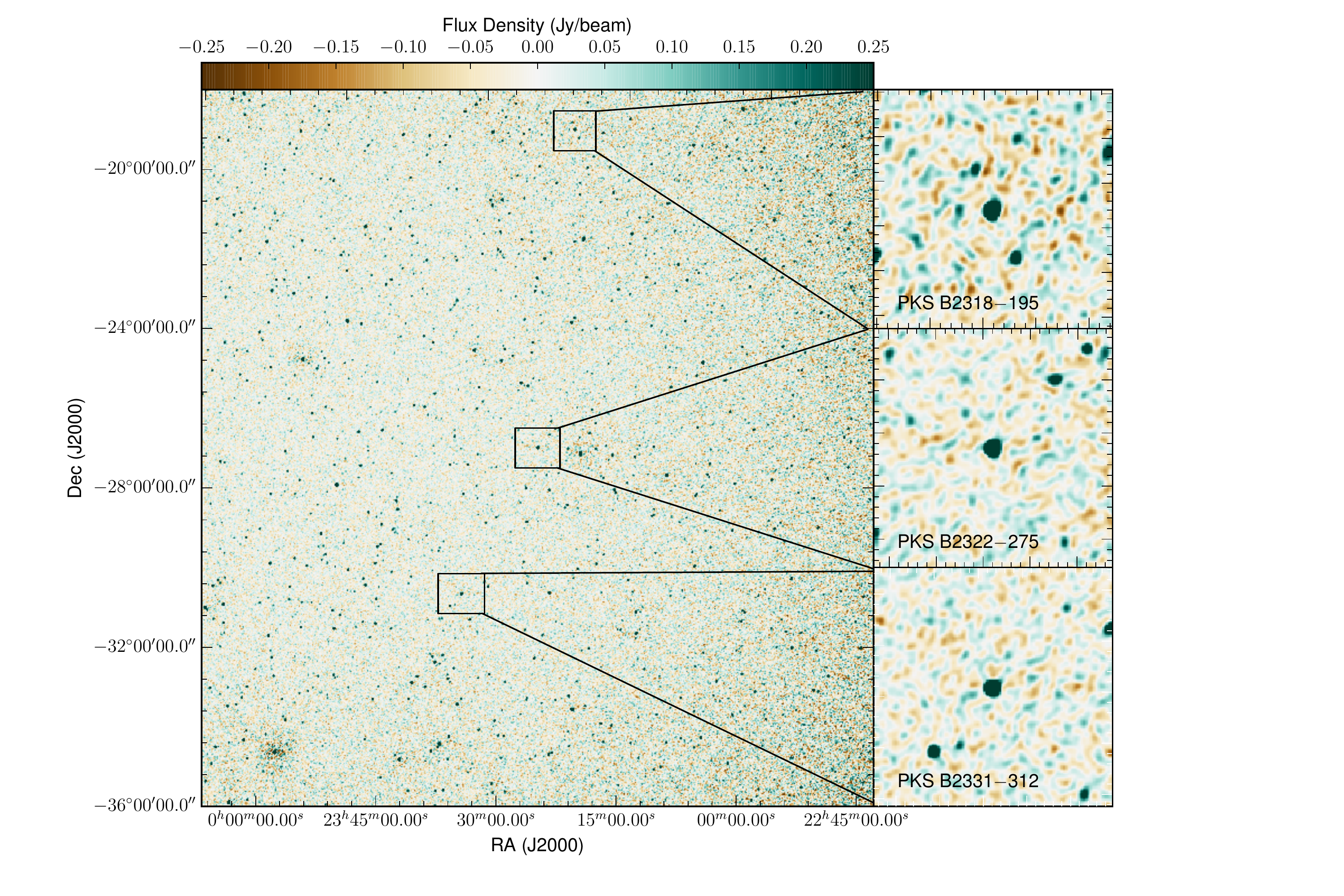}
\caption{MWA 155\,MHz image of the EoR0 field, centered on \pks.  The
  image is a single 2-min integration from 2014~November~06 and is roughly
  $17\degr$ on a side.  We show $1\,{\rm deg}^2$ boxes around \pks, \pksb, and
  \refsrc,
  with expanded views around these sources in the insets to the
  right.}
\label{fig:image}
\end{figure*}

Once the initial variability from \pks\ was identified, we searched
for variability from other objects in the field.  We analyzed data
from the MWA Real-Time System (RTS;
\citealt{2008ISTSP...2..707M,tin13}), which derives calibration
solutions including flux densities for hundreds of sources over the
MWA field-of-view during each integration.  For each of the sources
measured by the RTS (down to 155-MHz flux densities of about 1\,Jy/beam,
depending on the position in the MWA primary beam) we examined the
modulation during the observation.  Most of the sources had no
substantial excess variability (confirming the results of
\citealt{2014MNRAS.438..352B} and \citealt{2014PASA...31...45H}).  However, we did
identify a second source that had significant ($>30$\% over a 2-min
interval) modulation, which we identified as \pksb.  We focus on these
two sources; both are bright and have no nearby neighbors to corrupt
the flux density measurements (Fig.~\ref{fig:image}).

\subsection{Imaging Analysis}
To refine our analysis, we retrieved the observations from
2014~November~06 11:30 UT to 2014~November~06 14:14 UT from the MWA
archive.  The processing followed standard MWA procedures
\citep[e.g.,][]{2014PASA...31...45H}.
We calibrated the data using an observation of 3C~444, then rotated
the phases of the visibilities to center the field on \pks.  We imaged
individual 2\,s integrations in the XX and YY instrumental
polarizations using \texttt{WSClean} \citep{2014MNRAS.444..606O},
performing 200 \texttt{CLEAN} iterations on each and making images
with $3072\times 3072$ $0\farcm45$ pixels, with synthesized beam
full-width at half maximum of $2\farcm3$.  Finally, we corrected the
instrumental polarization to Stokes I (total intensity) using the
primary beam from \citet{2015RaSc...50...52S}.  For a single 2-s
integration, the average image noise was 0.16\,Jy/beam.  In
Figure~\ref{fig:image} we show a 2\,min integration with a slightly
wider field-of-view, which has a rms noise of about 0.05\,Jy/beam.

\begin{figure*}
  \plotone{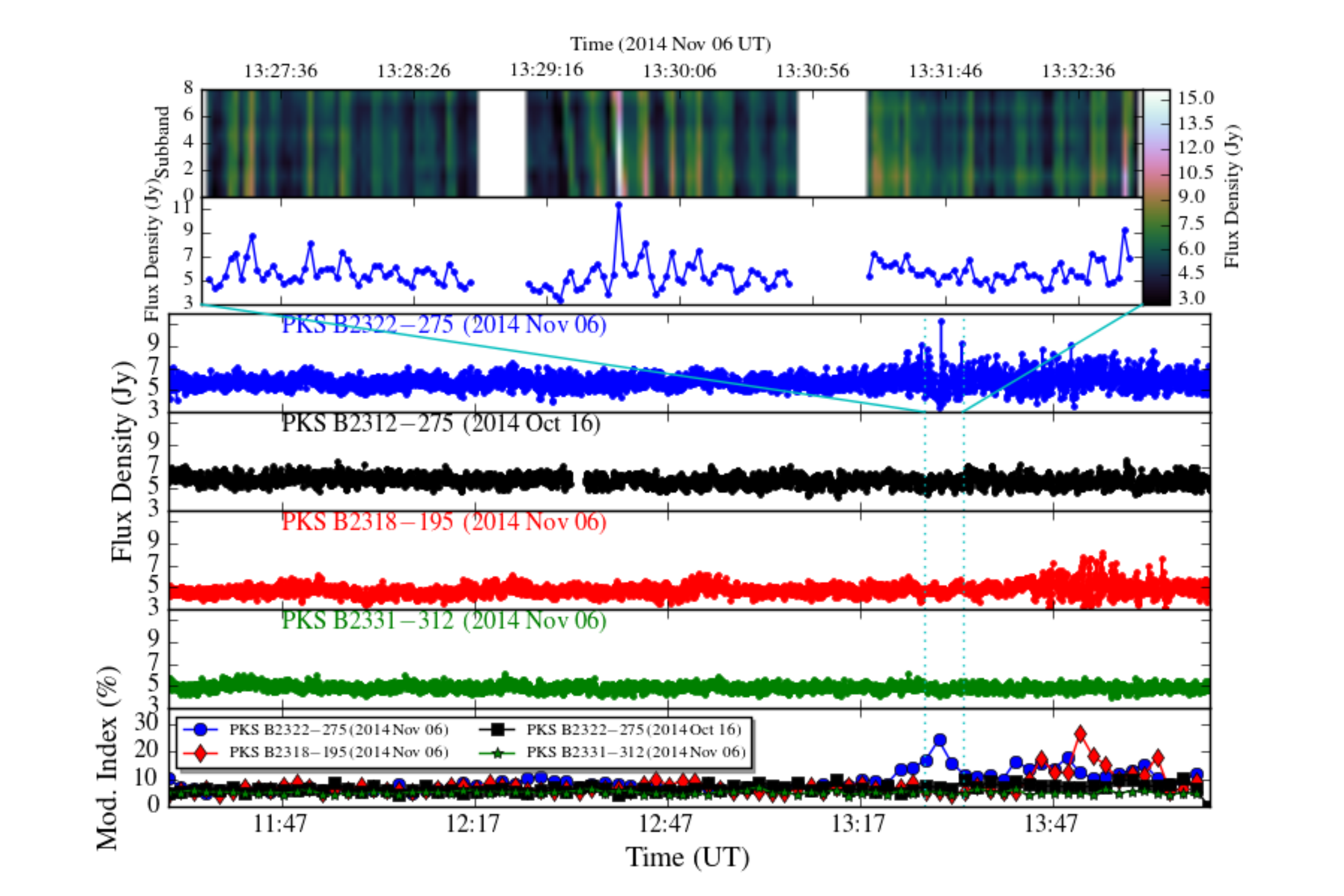}
\caption{Bottom section: flux densities of \pks\ on 2014~November~06
  (blue), \pks\ on 2014~October~06 (black), \pksb\ on 2014~November~06
  (red), and \refsrc\ on 2014~November~06 (green) measured with the
  MWA.  The flux densities were measured from individual 2\,s
  integrations, with uncertainty of about 0.2\,Jy.
Note that the
  y-axis ranges for all four panels are the same.  
Bottom panel: the
  modulation index (rms flux density variation over a 2\,min interval
  divided by the average flux density for that interval) for the same
  data-sets.  The expected uncertainty on the modulation index under
  the assumption of no extra variability is $<1$\%.  The top section shows
  the flux density of \pks\ along with the dynamic spectrum, measuring
  the flux density across 8 different sub-bands (3.84\,MHz each, with
  frequency increasing from the bottom) for
  the period indicated.  The gaps are from the intervals between
  subsequent 2-min observations.  }
\label{fig:flux}
\end{figure*}

We then used \texttt{Aegean} \citep{2012MNRAS.422.1812H} to measure
the flux densities and positions of \pks, \pksb, and a number of other
sources with comparable (4--10\,Jy) flux densities.  Both targets had
mean flux densities consistent with catalog values (5.8\,Jy at
160\,MHz for \pks, and 4.8\,Jy at 155\,MHz for \pksb;
\citealt{1995AuJPh..48..143S,2014PASA...31...45H}).  For each
integration we determined the mean fractional variation --- caused by
residual instrumental/ionospheric effects --- of a number of other
sources and divided it from each integration.  We plot the resulting
time series in Figure~\ref{fig:flux}.  We also show the modulation
index (rms divided by mean over a 2\,min interval) for \pks\ and
\pksb.  For comparison, we plot the flux densities of one of the
reference sources (\refsrc, also shown in Fig.~\ref{fig:image}) and of
\pks\ on another night with a comparable data-set (2014 October 16,
chosen randomly).  During the quiescent periods both \pks\ and
\pksb\ have modulation comparable to that for the reference source and
\pks\ on the comparison night (roughly 5\%, consistent with the
measured noise properties).  But at the peak, both \pks\ and
\pksb\ show modulation of $>100$\% peak-to-peak on 2-s timescales,
corresponding to rms variations of $>25$\% averaged over 2-min.  For
\pks\ we see that the brightest spike occurs first at the upper edge
of the bandpass before progressing down, giving large apparent swings
in the instantaneous spectral index.  The variations for \pks\ appear
roughly 20\,min before those in \pksb, which is about $9\degr$ north
of \pks\ and at a slightly higher solar elongation
($\epsilon=114\degr$ for \pks\ vs.\ $\epsilon=118\degr$ for \pksb)
appears to start 20--30\,min after that for \pks.  There is no
indication of increases in the amplitude of position fluctuations
during the period of most intense variability, with fluctuations of
about $5\arcsec$.

In Figure~\ref{fig:powspec} we show temporal power spectra for
\pks\ and \pksb\ (again with \refsrc\ and with \pks\ from another night for
comparison) created from the parts of Figure~\ref{fig:flux} with
strong variability. Both \pks\ and \pksb\ have considerably more power
at frequencies above roughly 0.06\,Hz than the comparison observation.
Below this point the power spectra are similar, suggesting that this
is the transition to ``normal'' variability (likely dominated by the
ionosphere, which should affect all sources roughly equally).  Above
the transition both scintillating sources are similar, declining with spectral slopes of roughly $-1$.

\section{Discussion}
\label{sec:discuss}
\subsection{Origin of Variability}
\label{sec:origin}
We can reject intrinsic variability since it appears for more than one
source at roughly the same time.  Moreover, light travel time
arguments for an AGN rule out the 2-s timescale we see, and the
extreme apparent spectral swings do not resemble any known source
behavior.  Instrumental effects are also ruled out, since the
timescales for instrumental changes (30\,min) are far longer than the
fluctuations we see and only two sources vary.  We are then left with
propagation effects, i.e., scintillation due to inhomogeneous,
turbulent plasma along the line of sight.  We must consider whether
scintillation is of interstellar, interplanetary, or ionospheric
origin.

Refractive interstellar scintillation causes much slower (minutes to
months) tens-of-percent variations in the flux densities of compact
radio sources, with the timescale increasing at lower frequencies
\citep[e.g.,][]{1986ApJ...307..564R}, hence can be clearly ruled out.
Diffractive scintillation can be substantially faster
\citep{1998ApJ...507..846C}, but would require sizes that imply
brightness temperatures of $\gg 10^{12}$\,K --- impossible for an AGN
\citep{1994ApJ...426...51R}.

{Ionospheric scintillation affects sources with angular sizes
  $\lesssim 10\arcmin$ \citep{loi15b}, causing joint position- and
  amplitude fluctuations on timescales of $\gtrsim 20\,$s
  \citep{2001isra.book.....T}, with the former a significant source of
  position wander in MWA data \citep{2014PASA...31...45H,loi15b} and
  the latter usually not very significant (although
  see \citealt{2014JGRA..11910544F}).  Therefore, ionospheric
  refraction should affect the 
  thousands of unresolved sources in each image.  In contrast, here the very strong
  amplitude fluctuations without any position shifts for unresolved sources
  and the rather small number of scintillating  sources argue
  strongly against scattering by ionospheric plasma.  }

\begin{figure*}
 \plottwo{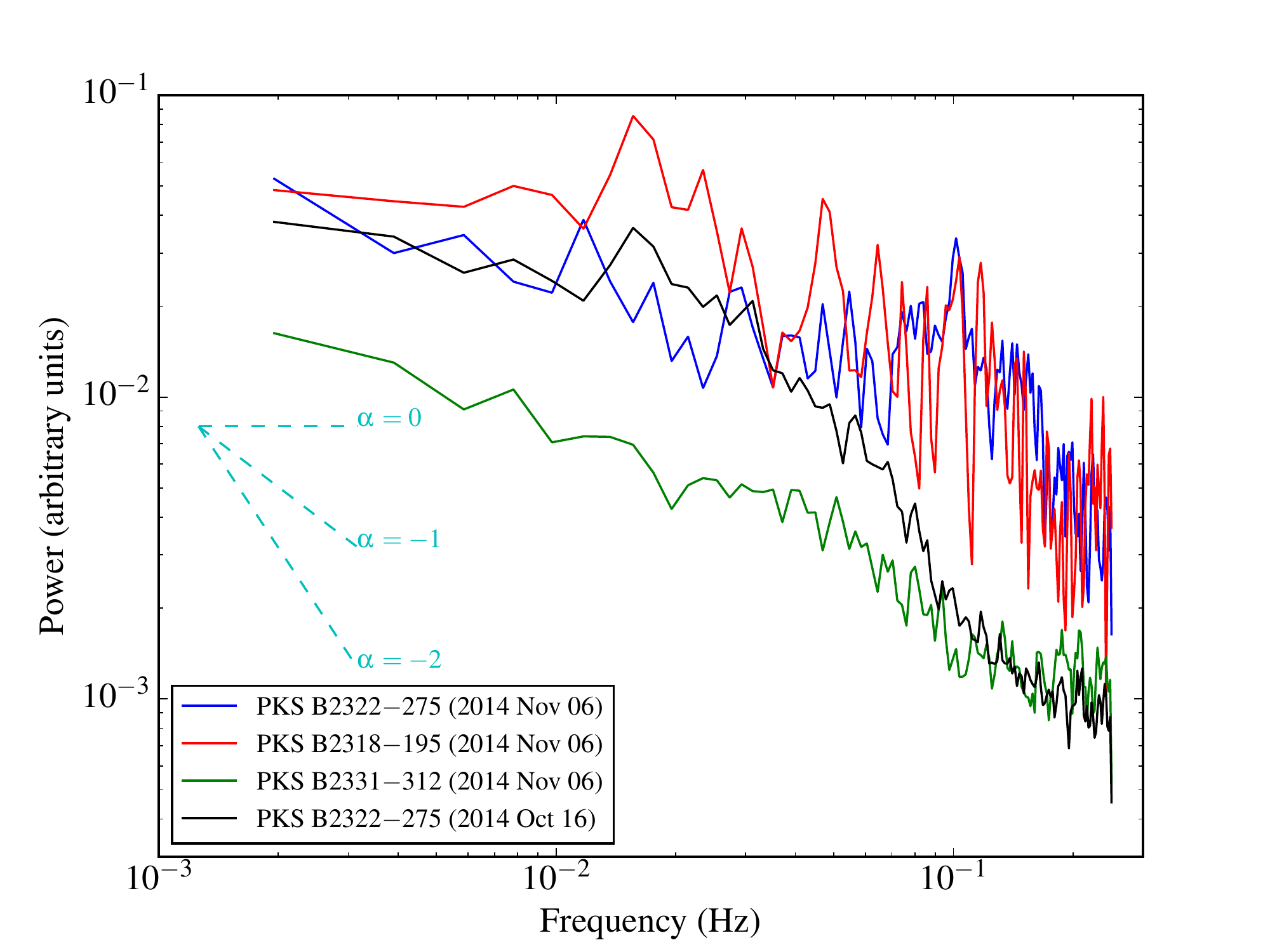}{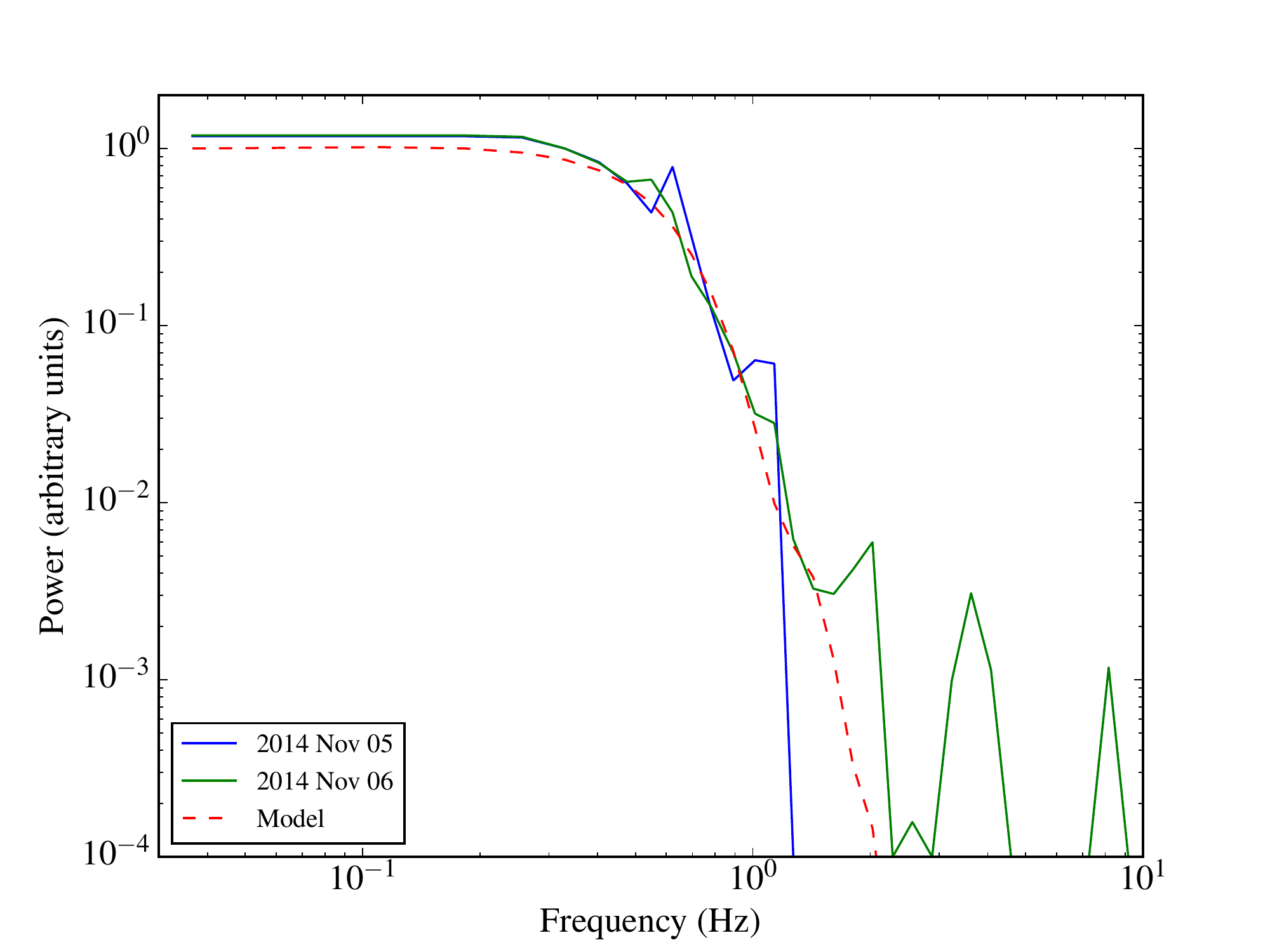}
\caption{Left: temporal power spectra of the flux densities for
  \pks\ (blue, with 62\,min of data), \pksb\ (red, with 37\,min of
  data), and \refsrc\ (green, with 235\,min of data), using the data
  from Figure~\ref{fig:flux} when large-amplitude scintillations are
  evident. The power spectra are estimated using Welch's method,
  averaging spectra with 256 data-points and 50\% overlap.  We also
  show the power spectrum of \pks\ from our comparison night (black).
  The cyan dashed lines show power-laws with slopes $\alpha=0$, $-1$,
  and $-2$ as labeled.  Right: temporal power spectra of the flux
  density for PKS B2211$-$388, measured by the Ooty radio telescope at
  327-MHz.  The power spectrum for 2014~November~05 (blue) shows
  substantially less power above 1\,Hz than that for 2014~November~06
  (green; note that both power spectra have been normalized below
  0.1\,Hz).  The model fit (red dashed line) is for a solar wind
  velocity of $\approx 500\,\kms$.  }
\label{fig:powspec}
\end{figure*}

In contrast, IPS requires significant source structure on scales
$<1\arcsec$ \citep{1964Natur.203.1214H,1968MNRAS.138..393L}.
All three sources discussed here are regularly monitored
for IPS with Ooty: both \pks\ and \pksb\ are often seen to scintillate
with Ooty, consistent with angular sizes of $\lesssim 0\farcs1$ (at
327\,MHz), while Ooty sees very little (if any) scintillation toward
\refsrc.    Note that Ooty does not observe past $\epsilon=100\degr$,
so it could not monitor \pks\ during this event, although it did
observe a number of nearby sources at $\epsilon>90\degr$
(Figs.~\ref{fig:powspec} and \ref{fig:ooty}) that showed enhanced
scintillation on 2014~November~06 at frequencies $>2\,$Hz.

Moreover, the two scintillating sources are detected  
at frequencies up to 20\,GHz \citep{2010MNRAS.402.2403M}, suggesting
the presence of compact structure\footnote{\pks\ is one of only two
  sources in this field brighter than 5\,Jy at 180\,MHz which is
  unresolved at 20\,GHz (determined by its 6\,km-baseline flux
  density).  \pksb\ is partially resolved at 20\,GHz, suggesting some
  arcsecond-scale structure but also significant compact emission.}
\citep{2013MNRAS.434..956C}.  In contrast, 
\refsrc\ has a similar flux density at 155\,MHz
\citep{2014PASA...31...45H} but is not detected above 5\,GHz (implying
a steeper spectral index $\alpha\approx-1.2$, with $S_\nu\propto
\nu^\alpha$) and hence likely has more extended emission.
This can explain why \pks\ and \pksb\ were
observed to be the only significant bright scintillators.  That would
not be the case for an ionospheric origin.

Finally, NASA's Omniweb database\footnote{See
  \url{http://omniweb.gsfc.nasa.gov/}.} suggests that only weak
auroral and ionospheric activity occurred on both nights, with no
clear CME arriving at Earth, modest $K_p$ indices\footnote{See
  \url{http://www.ngdc.noaa.gov/stp/GEOMAG/kp\_ap.html}.} (global
average magnetic activity) of about 2, and temporal behavior over the
whole field typical of quiet ionospheric conditions\footnote{While we
  do see a modest enhancement in the auroral activity (given by the AE
  index) that peaks near 13:25\,UT, the enhancement is not very
  notable, there are multiple enhancements that day, and there are
  even more significant enhancements on our comparison night which are
  not coincident with scintillation.}  (S.~T.~Loi 2015, pers.~comm.).
Absent any other viable explanation for what we see, and since what we
see agrees with all of the known properties (amplitude, time-scale,
and source selection) of IPS (as in \citealt{1964Natur.203.1214H}), we
conclude IPS the most likely origin of the anomalous variability.
This is largely confirmed by the Ooty Radio Telescope observations.

\subsection{Variability Properties}
The modulation of \pksb\ appears to start 20--30\,min after that for
\pks, and that the period of enhanced modulation lasts for about
30\,min for both sources (although some of the modulation may extend past the
end of our observing period).  {Assuming the modulation is
  caused by IPS from a CME}, the elongation difference implies a
speed of $\sim 500\,{\rm km\,s}^{-1}$ for a distance of 0.1\,AU. This
speed is consistent with those inferred from Ooty observations on
2014~November~06, where a number of sources at elongations $>90\degr$
that were in the same direction as \pks\ and \pksb\ showed
significantly enhanced scintillation (see Fig.~\ref{fig:powspec} for
an example).  For
a solar wind speed of $500\,v_{500}\kms$, we infer a minimum
transverse size of the CME of ${9} \times 10^{5}v_{500}\,$km from the
duration of the modulation.  However, given typical radial expansion
of CMEs \citep[e.g.,][]{2000ApJ...530.1061M} we expect radial sizes of
0.2--0.4\,AU.  Therefore we believe that the duration of the IPS is
set by the thickness of {some sub-structure within the putative
  CME.}

As there is significant variability down to our sampling of 2\,s, we
infer CME structure on scales down to $\lesssim$1000$v_{500}$\,km.  We
see an apparent transition from a flat power spectrum to a decline in
Figure~\ref{fig:powspec}.  This may be associated with the Fresnel
scale, above which IPS is suppressed \citep{1990MNRAS.244..691M}.
{However, actually measuring the Fresnel scale and
directly constraining the turbulent properties generally
requires considerably faster sampling.}
Instead, since the rms fluctuations are $<100$\%, it is likely that the sources
are partially resolved by IPS,  the change in spectral slope could
have more to
do more with intrinsic source sub-structure (this would be consistent
with the change in scintillation with radio frequency in
Fig.~\ref{fig:flux}).  

\subsection{Identification of Candidate CMEs}
\begin{figure*}
  \plotone{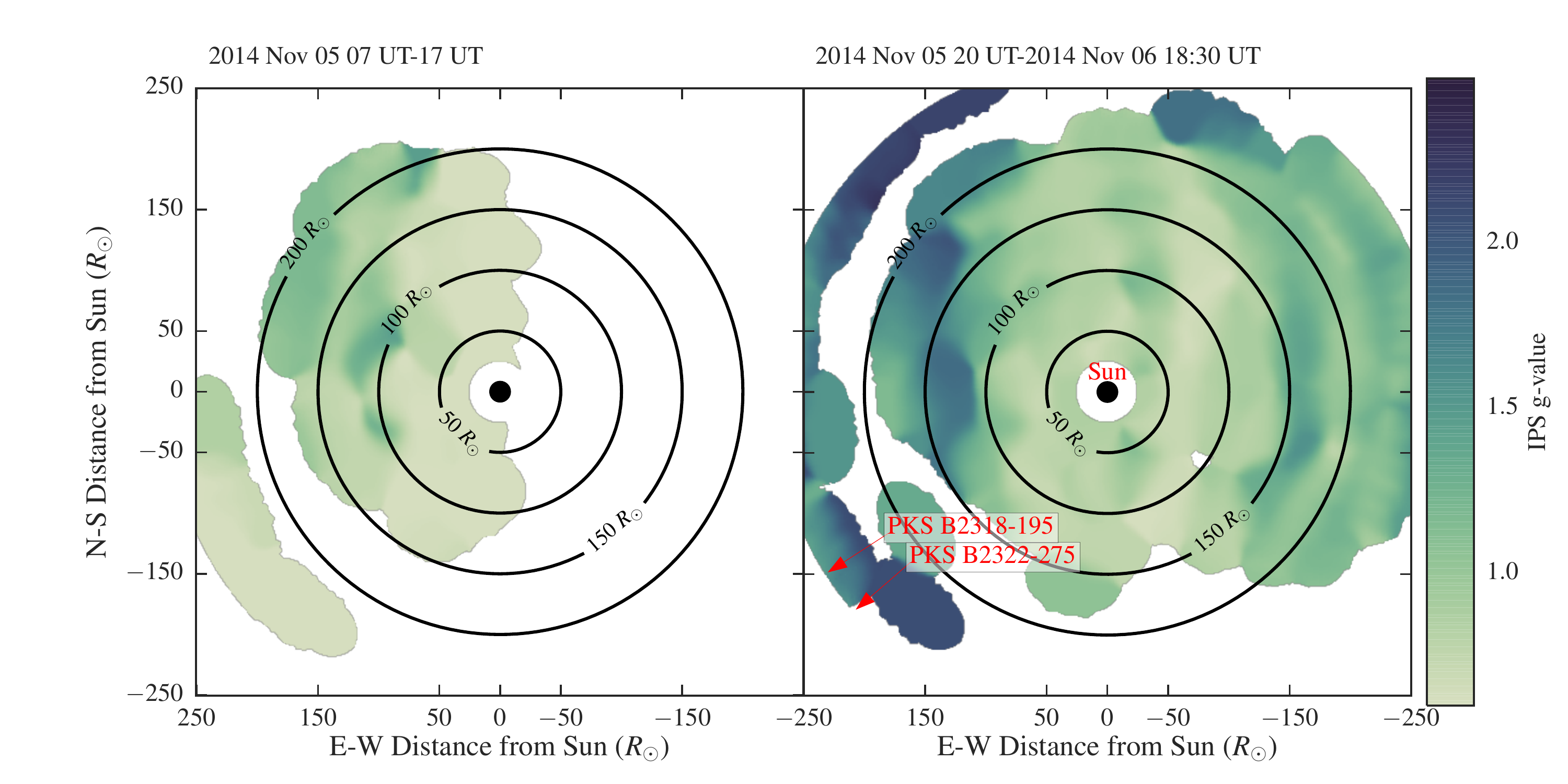}
\caption{Ooty radio telescope 327-MHz scintillation images.  Left:
  image for the
  period 2014~November~05 07~UT to 17~UT.  Right: image for the period 
  2014~November~05 20~UT to 2014~November~06 18:30~UT (which
  spans our scintillation event at 13:30\,UT).  The images show the
  level of scintillation measured by the $g$ parameter
  \citep[e.g.,][]{2010SoPh..265..137M}, where $g=1$ is a normal level
  of turbulence.  Radial distance from the center shows the derived
  heliocentric distance of the scattering medium: concentric circles
  with radii of 50, 100, 150, and 200\,$R_{\odot}$ are plotted and
  labeled.  North is up, and East to the left.  The arrows show the
  directions toward \pks\ and \pksb\ starting at a distance of 1\,AU
  (215\,$R_\odot$).  The image from 05~November shows very little IPS,
  while the image from 06~November shows a significant increase in IPS
  toward the South East where the MWA data are.
}
\label{fig:ooty}
\end{figure*}

At the time of the observation, \pks\ was at $\epsilon=114\degr$ and a
position angle of about $114\degr$ (east of north).  So any plasma
structure that caused the scintillation would have to have been
outside the orbit of the Earth and below the ecliptic
(Fig.~\ref{fig:ooty}).  With typical speeds of 500--1000\,\kms, we
searched for CMEs that were observed to occur around 3--5 days before
our observations. Among the CMEs identified by Computer Aided CME
Tracking software (CACTus; \citealt*{2009ApJ...691.1222R}) we
identified a number with reasonable ejection times and angles. The
scintillation image produced by the Ooty Radio Telescope
(Fig.~\ref{fig:ooty}) for 2014~November~05 up to 17:00\,UT shows
little scintillation in the direction towards \pks, while the
scintillation increased significantly after 20:00\,UT and contiuing on
2014~November~06.  Without more detailed tracking or simulation of
heliospheric structures, especially beyond Earth, we cannot
conclusively determine whether a specific CME or region resulting from
multiple CME interactions caused the observed scintillations.

\section{Conclusions}
\label{sec:conc}
Plans for observations of IPS with the new generation of low-frequency
radio telescopes, namely LOFAR and the MWA, concentrate on dedicated
high-time-resolution ($\lesssim 10\,$ms) observations of pre-selected
sources.  The temporal and frequency dependence of IPS flux variations
help probe the structure of the solar wind (including possible CMEs)
\citep[e.g.,][]{2010SoPh..265..293O,2013SoPh..285..127F,bow13}.  These
observations are similar in strategy to dedicated programs with
existing facilities like the Ooty radio telescope
\citep{2010SoPh..265..137M}, which measures scintillation of 400--800
sources each day distributed over $\epsilon<100\degr$.

{Assuming the variability we see is related to IPS,} we have
discussed a different regime.  First, we used commensal night-time
observations with the MWA, so $\epsilon>90\degr$.
Second, the sources observed at facilities like Ooty correspond to
155\,MHz flux densities $S_{155}\gtrsim 10\,$Jy, which have an areal
density of $\approx 100\,{\rm sr}^{-1}=0.03\,{\rm deg}^{-2}$
\citep{2012ApJ...755...47W}.  Over the 600\,$\deg^2$ MWA field-of-view
that means we would detect at least 20 of these bright sources.  In
fact, we can sensitively probe flux density variations of 20\% in a
single 2-s integration for sources as weak as 1\,Jy, which have a sky
density $30\times$ higher.  Therefore, the density of sight-lines on the sky
available to probe IPS is a factor of five finer, and in fact weaker
sources are observed to vary (at lower confidence) along with the
sources discussed here.  While the 2\,s integrations used here average
over much of the higher-time-resolution IPS structure, quasi-real time
analysis (as in \citealt{2008ISTSP...2..707M}) can be used to identify
scintillating sources which can then be followed at higher time
resolution using dedicated observations \citep{tre15}.

One difficulty in interpreting our results, though,
comes from the fact that we are looking outside the Earth's orbit,
where the geometry does not allow easy identification  of the
distances to 
turbulent structures \citep[cf.,][]{1978ApJ...220..346A}.
Even so, we can still examine the statistics of IPS-like variability
over multi-year timescales, correlating it with dedicated ground- and
space-based solar observing.  We can also probe the fraction of
compact ($<$arcsec) emission from the roughly $10^4$ bright,
high-latitude (where scatter broadening will not matter;
\citealt{ne2001}) sources to be observed in various MWA surveys
(cf.\ hundreds of sources in
\citealt{1968MNRAS.138..393L,2000AA...362...27J}, although see
\citealt{1990MNRAS.247..491H}), providing a useful adjunct to
calibrator surveys and allowing connections between spectral and
spatial properties of sources.

Finally, we note that the existence of strong IPS in nighttime
observations of a primary field for the MWA EoR experiment using
nominal imaging parameters may be concerning.  However, a detailed
analysis by \citet{tt15} shows that under all
but extreme conditions, IPS will not affect the detectability of the
EoR signal.

\acknowledgments

We thank an anonymous referee for a thorough and thoughtful review.  We
thank B.~Jackson, C.~Loi, and A.~Rowlinson for helpful conversations.
This scientific work makes use of the Murchison Radio-astronomy
Observatory, operated by CSIRO. We acknowledge the Wajarri Yamatji
people as the traditional owners of the Observatory site. Support for
the MWA comes from the U.S. National Science Foundation (grants
AST-0457585, PHY-0835713, CAREER-0847753, and AST-0908884), the
Australian Research Council (LIEF grants LE0775621 and LE0882938), the
U.S. Air Force Office of Scientific Research (grant FA9550-0510247),
and the Centre for All-sky Astrophysics (an Australian Research
Council Centre of Excellence funded by grant CE110001020). DLK is
additionally supported by NSF grant AST-1412421.  PKM acknowledges
partial support from ISRO.  Support is also provided by the
Smithsonian Astrophysical Observatory, the MIT School of Science, the
Raman Research Institute, the Australian National University, and the
Victoria University of Wellington (via grant MED-E1799 from the New
Zealand Ministry of Economic Development and an IBM Shared University
Research Grant). The Australian Federal government provides additional
support via the Commonwealth Scientific and Industrial Research
Organisation (CSIRO), National Collaborative Research Infrastructure
Strategy, Education Investment Fund, and the Australia India Strategic
Research Fund, and Astronomy Australia Limited, under contract to
Curtin University. We acknowledge the iVEC Petabyte Data Store, the
Initiative in Innovative Computing and the CUDA Center for Excellence
sponsored by NVIDIA at Harvard University, and the International
Centre for Radio Astronomy Research (ICRAR), a Joint Venture of Curtin
University and The University of Western Australia, funded by the
Western Australian State government.  This research made use of APLpy,
an open-source plotting package for Python hosted at
\url{http://aplpy.github.com}.

{\it Facilities:} \facility{Murchison Widefield Array}, \facility{Ooty
  Radio Telescope}.

\bibliographystyle{apj} 
%\bibliography{ips_short}

\end{document}